\documentclass[prl, reprint]{revtex4-1} 

\usepackage{amsmath} 

\usepackage{graphicx} 

\usepackage{color} 

\usepackage{amsmath}
\usepackage{nicefrac}
\usepackage{graphicx}
\usepackage{wrapfig}
\usepackage{setspace}

\usepackage{url}

\usepackage{hyperref}

\newcommand{\djs}{\mathcal{D}_{\rm{JS}}}

\newcommand{\clobs}{C_l^{obs}}

\newcommand{\rl}{\frac{1}{2}(C_l^{obs}+C_l)}

\begin{document}

\title{Informational approach to cosmological parameter estimation}

\author{Michelle Stephens}
\email{Michelle.M.Stephens.GR@dartmouth.edu}
\affiliation{Department of Physics and Astronomy\\ Dartmouth College, Hanover, New Hampshire 03755, USA}

\author{Sara Vannah}
\email{Sara.A.Vannah.GR@dartmouth.edu}
\affiliation{Department of Physics and Astronomy\\ Dartmouth College, Hanover, New Hampshire 03755, USA}

\author{Marcelo Gleiser}
\email{Marcelo.Gleiser@dartmouth.edu}
\affiliation{Department of Physics and Astronomy\\ Dartmouth College, Hanover, New Hampshire 03755, USA}

\date{\today}

\begin{abstract}
We introduce a new approach for cosmological parameter estimation based on the information-theoretical Jensen-Shannon divergence (${\cal D}_{\rm JS}$), calculating it for models in the restricted parameter space $\{H_0, w_0, w_a\}$, where $H_0$  is the value of the Hubble constant today, and $w_0$ and $w_a$ are dark energy parameters, with the other parameters held fixed at their best-fit values from the Planck 2018 data. As an application, we investigate the $H_0$ tension between the Planck temperature power spectrum data (TT) and the local astronomical data by comparing the $\Lambda$CDM model with the $w$CDM and the $w_0w_a$CDM dynamic dark energy models. We find agreement with other works using the standard Bayesian inference for parameter estimation; in addition, we show that while the ${\cal D}_{\rm JS}$ is equally minimized for both values of $H_0$ along the $(w_0,w_a)$ plane, the lines of degeneracy are different for each value of $H_0$. This allows for distinguishing between the two, once the value of either $w_0$ or $w_a$ is known.
\end{abstract}

\maketitle

\section{Introduction}\label{sec:Intro}
The concordance cosmological model, $\Lambda$CDM, is in excellent agreement with data spanning a broad range of redshifts, including the temperature anisotropies of the cosmic microwave background (CMB) \cite{Akrami:2018}, the large-scale galaxy clustering feature of baryon acoustic oscillations (BAO) \cite{Sanchez:2012}, and the luminosity-redshift relation of local sources, calibrated primarily with type-Ia supernovae data (SNeIa) \cite{Lopez-Corredoira:2016}. The model has six independent parameters, assuming the dark energy (DE) equation of state is a constant $w = -1$, and a flat universe: the amplitude and spectral index of the primordial density perturbations, $A_s$ and $n_s$, respectively; the reionization optical depth $\tau$; the present-day Hubble parameter $H_0$; and the present-day physical baryon and dark matter densities $\Omega_b h^2$ and $\Omega_c h^2$, respectively, where $h = H_0/100$. 

The Friedmann equation for a spatially flat universe with a cosmological constant $\Lambda$, in the matter-dominated era ($z \ll 3200$) can be written as $\Omega_b + \Omega_c + \Omega_{\Lambda} = 1$, where $H_0$ determines the critical density normalization on $\Omega_X$. Thus, the cosmic expansion history and structure formation in the universe is sensitive to the relative contributions of $\Omega_m = \Omega_b + \Omega_c$, and DE. Despite the overall success of $\Lambda$CDM, statistically significant tensions exist between early-universe parameter inference and direct local measurement, most notably in the value of the Hubble parameter today, $H_0$ \cite{Freedman:2017,Freedman:2019,Bernal:2016}. Recent results  indicate that the discrepancy does not appear to be dependent on the use of any one method of measuring $H_0$ in the late universe, yielding a persistent tension with early-universe measurements between $4.0\sigma$ and $5.8\sigma$ 
\cite{Verde:2019}.

Measurements of the CMB anisotropies at $z \approx 1100$ by the Planck \cite{Aghanim:2018} and WMAP \cite{WMAP9} missions constrain the combinations $\Omega_b h^2$ and $\Omega_c h^2$, but degeneracies prevent constraints of $H_0$ alone \cite{Zaldarriaga:1997, Efstathiou:1998, Howlett}. Local measurements can probe $H(z)$ directly through the luminosity-redshift relation, but distances to sources must be carefully calibrated to avoid systematic error. Uncertainties have been reduced to the subpercent level in the case of the Planck analysis, and to the 1\% level with recent advances in the local distance-ladder determinations \cite{Riess:2016, Riess:2019}. Excluding an as-yet unknown source of error in either of these analyses, the discrepancy may point to new early-universe physics beyond the standard cosmological model. 

Several possible resolutions to the Hubble tension have been proposed, including evolving DE with a phantom-like equation of state \cite{DiValentino:2017}, additional neutrinos \cite{PoulinNeff:2018, Carneiro:2018}, local voids \cite{Marra:2013}, and prerecombination modifications to DE (early dark energy) \cite{PoulinEDE:2018}, among many others \cite{DiValentinoVPT:2017,DiValentinoIDE:2017,Bolejko:2017,PoulinNeff:2018}. Given the many data sets, extending the cosmological parameter space and performing a Markov chain Monte Carlo (MCMC) analysis to determine the most likely parameters is a computationally expensive problem \cite{DiValentino:2015} and involves many complications in constructing the likelihood function arising from particular instrumentation, data-set considerations, and prior choices \cite{Ade:2013,Gerbino:2019, Smith:2020}.

Here, we propose an alternative approach to cosmological parameter estimation based on a measure from information theory known as the Jensen-Shannon divergence (${\cal D}_{\rm JS}$). We apply it to a one-parameter extension of the $\Lambda$CDM model, the $(w_0, w_a)$ parametrization of an evolving dark energy component. In  \hyperref[sec:ParameterDetermination]{Sec. II}, we review the standard maximum likelihood method before introducing the ${\cal D}_{\rm JS}$ and the information theory needed for its interpretation, with a toy example. In \hyperref[sec:Methods]{Section III}, we provide motivation for using the ${\cal D}_{\rm JS}$ to examine linearly evolving DE models and detail the numerical approach. Results are presented in \hyperref[sec:Results]{Sec. IV}. In \hyperref[sec:Conclusions]{Sec. V}, we highlight prospects for extending this study in future work.

\section{Parameter Estimation}
\label{sec:ParameterDetermination}
\subsection{Maximum likelihood estimation}\label{subsec:MaxLikelihood}
The standard approach for cosmological parameter estimation is a problem of Bayesian inference: We begin with a data set $D$, which we wish to accurately represent with a model parametrized by $\theta$. We assume a prior distribution over the parameters, $p(\theta)$. The prior ideally represents our best knowledge of the parameters, but in practice, it is commonly taken to be uniform. The model is specified by the form of the likelihood function, $\mathcal{L}(\theta) \equiv p(D | \theta)$---the probability that the data are observed, given the model. The posterior probability $p(\theta | D)$ is, by Bayes' rule, proportional to $p(D | \theta) p(\theta)$. The best parameters are inferred by maximizing the likelihood function. 

Details of this method applied to cosmology can be found in Refs. \cite{Trotta:2008, Bond:1997, White:1994}. The MCMC code \textsc{CosmoMC} \cite{Lewis:2002} can incorporate the likelihoods for various data sets as well as prior specifications; it is well validated and the most commonly used code for cosmological parameter estimation. In this work, we allow $w_0$ and $w_a$ to vary while setting $H_0$ to its value  first from early-universe parameter inference and then from local direct measurement. In future work we will extend the analysis using ${\cal D}_{\rm JS}$ to the full parameter space. Parameter estimates with confidence intervals can then be directly compared to maximum likelihood results.

\subsection{Jensen-Shannon divergence}\label{subsec:Information}
Shannon's seminal paper \cite{Shannon} provides the foundation for the definition and interpretation of the ${\cal D}_{\rm JS}$. We typically think of describing a message or event in terms of a distinct encoding scheme---a set of symbols $\mathcal{N} = {n_1, n_2, \ldots , n_L}$. For instance, in English we encode words with the $26$ letters of the alphabet, and full messages with additional characters for punctuation. The information content of a particular symbol is $I(n) = -\log_2 p(n)$, where $p(n)$ is the probability distribution over symbols in $\mathcal{N}$ determined from some collection of events encoded by $\mathcal{N}$.  The expected value of information in a particular event is then

\begin{equation}
\langle I \rangle = - \sum_{n \in \mathcal{N}} p(n) \: \log_2 p(n).
\end{equation}

The ${\cal D}_{\rm JS}$ is a measure of the difference between two probability distributions, based on the Kullback-Leibler divergence (DKL) between two distributions $p(n)$ and $q(n)$, defined as \cite{KullbackLeibler}

\begin{equation}
\mathcal{D}_\mathrm{KL}(p\:||\:q) = \sum_n p_n \log \left(\frac{p_n}{q_n}\right).
\end{equation}

If $q$ is treated as a model for some ``true" distribution $p$, DKL is a measure of the information lost in using $q$ rather than $p$. DKL is positive-definite, and it is zero only if the two distributions are the same (known as the identity of indiscernables). However, it is not symmetric and does not satisfy the triangle inequality: while we picture it as a ``distance'' between two distributions, it is not a metric. The ${\cal D}_{\rm JS}$ is a symmetrized extension of DKL that can be treated as a true metric on the space of probability distributions \cite{Lin:1991}. It is defined as 

\begin{equation}\label{eq:JSD}
\mathcal{D}_\mathrm{JS} = \frac{1}{2}\mathcal{D}_\mathrm{KL}(p\:||\:r) + \frac{1}{2}\mathcal{D}_\mathrm{KL}(q\:||\:r),
\end{equation}

where $r = \frac{1}{2}(p+q)$. Here, $0 \leq \mathcal{D}_\mathrm{JS} \leq 1$ if the logarithm used in the DKL is base 2. In this case, information is measured in bits. With the exception of the next example, we use the natural logarithm, so that $0 \leq \mathcal{D}_\mathrm{JS} \leq \ln(2)$.

As a simple illustration of the ${\cal D}_{\rm JS}$, consider a collection of short messages in English. Suppose we eliminate the spaces, capitalization, punctuation, etc., so that the set of symbols from which each message is drawn is simply the $26$-character English alphabet. Using the ${\cal D}_{\rm JS}$, we can find out how well each of these messages models the phrase ``Cosmology Rocks."

\begin{figure}[h!]
\includegraphics[width=\linewidth]{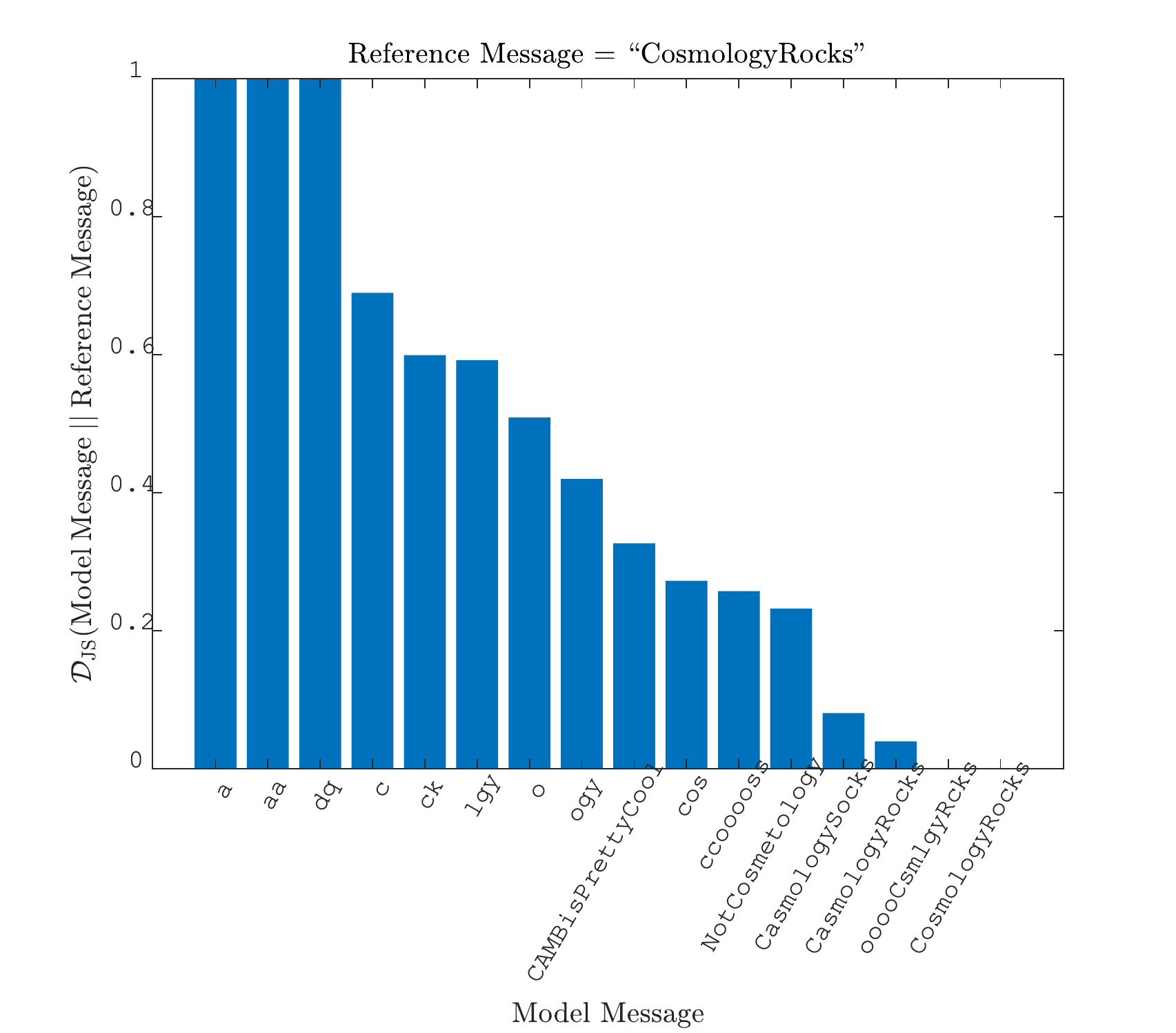}
\caption{${\cal D}_{\rm JS}$ between the model messages and the reference message is shown, where the messages are case insensitive and drawn from the English alphabet of $26$ letters. This illustrates that the ${\cal D}_{\rm JS}$ is sensitive only to the identity and relative frequencies of letters in a message. It is maximal for distributions with nothing in common and minimal (at zero) for identical distributions.}
\label{fig:JSD_English}
\end{figure}

It is clear that only the relative \textit{frequency} of letters in each of the phrases determines the ${\cal D}_{\rm JS}$ to the reference message---variables such as ordering of the letters and overall length are irrelevant. Model messages in which none of the letters appears in the reference message are maximally divergent, and those which letter frequencies close to the reference message have low divergence. These results are shown in \hyperref[fig:JSD_English]{Fig. 1}.

The more interesting examples come from comparing the ${\cal D}_{\rm JS}$ for model messages like ``lgy," ``cos," and ``ccooooss." In the first model, the entire message appears in the reference with the appropriate letter frequency. In the second, the model is comprised of the three most common letters from the reference but with incorrect frequencies. In the final model, the most common letters from the reference appear with the correct frequency. We conclude, then, that the ${\cal D}_{\rm JS}$ is sensitive to the most salient features of a given distribution. 

It is this property that makes it a good candidate for examining the divergence between the measured angular power spectrum of the microwave background and a model's prediction for it. If we replace the alphabet with the multipole moment $l$, and the frequency with $C_l$, it is the location and relative scaling of the acoustic peaks that provides the bulk of the CMB's sensitivity to cosmological parameters. Finally, we note that if $q_n = p_n \pm \delta p_n$, with $\delta p_n \ll p_n$, expanding the ${\cal D}_{\rm JS}$ to first order in $\delta p_n$ gives a measure proportional to the chi square.
\vspace{-0.5cm}
\section{Methods}\label{sec:Methods}

\subsection{Dark energy and the $H_0$ tension}\label{subsec:DarkEnergy}

The $H_0$ tension can be stated in this way: late-time scale factor expansion is occurring faster than we would expect from $\Lambda$CDM, with parameter constraints inferred from early-universe data. Framed this way, it is easy to see why most of the proposed resolutions involve modifying DE in some way. Prerecombination modifications to DE can alter the sound horizon $r_s$ and thus change the inferred $H_0$, while minute shifts in other parameters maintain the agreement with CMB anisotropies \cite{PoulinEDE:2018}. Late-time modifications are an obvious mechanism to alter the expansion history and galaxy clustering, but are constrained by other measurements, notably BAO \cite{Aubourg:2015, Haridasu:2017, Cheng:2014}. Constraining the DE equation of state is challenging, because density parameters and $H(z)$ are sensitive to a function of its integral over redshift. 

Observational surveys like the DESI probe \cite{Vargas-Magana:2019} will be able to provide direct constraints on $w(z)$. Until then, phenomenological models have been introduced to capture what the general behavior of $w \neq -1$ might look like, and its influence on cosmological observables. A common parametrization for evolving DE is the linear evolution model $w = w_0 + w_a(1-a)$, where $w_0$ is the value today and $w_a = -dw/da$ \cite{Chevallier:2000}. In this framework, $\Lambda$CDM corresponds to $w_0 = -1, w_a = 0$, and other constant-$w$ models can be considered by setting $w_a = 0$. 

Several studies have extended the $\Lambda$CDM basic six-parameter model to include these parameters, constraining them in the extended space via standard MCMC max-likelihood methods. References \cite{Freedman:2010,DiValentino:2015,DiValentino:2017,Xia:2013} are an incomplete list.
\vspace{-0.25cm}
\subsection{Data and numerical approach}\label{subsec:Numerics}
We compare a model's prediction for the angular power spectrum to the Planck 2018 data by computing $\mathcal{D}_\mathrm{JS}\left(F_l^\mathrm{mod} \:||\:F_l^\mathrm{plk}\right)$ as in \hyperref[eq:JSD]{Eq. (3)}, where $F_l^\mathrm{mod}$ and $F_l^\mathrm{plk}$ are determined from the model-predicted and Planck data-calculated angular power spectra, respectively. That is,

\begin{equation}
F_l = \frac{D_l}{\sum\limits_l D_l},
\end{equation}

where $D_l = l(l+1)C_l/2\pi$. Here, ${F_l}$, which we call the modal fraction (see Refs. \cite{Gleiser:2012,Stephens:2018} for more details), measures the relative probability for a given angular mode over the full data set, playing a similar role to the probability of occurrence in a message of a letter belonging to a given alphabet in Shannon's information entropy. Each data set is normalized to unity so that we can interpret $F_l$ as a probability distribution.
We use $D_l$ to compute the ${\cal D}_{\rm JS}$ since it more clearly distinguishes the acoustic features.

The unbinned $C_l$ computed from the temperature fluctuations observed by Planck can be found on the Planck Legacy Archive \cite{PLA}; we use the third public release of the baseline high-$l$ Planck TT power spectrum. (In future work, we plan to include polarization and temperature-polarization cross-correlation power spectra.) The cosmological Boltzmann code \textsc{CAMB}   \cite{LewisChallinor} is used to compute the angular power spectrum for a given model. The base set of cosmological parameters and their best-fit values as determined by the Planck 2018 analysis are summarized in \autoref{tab:Params}. All parameters except $H_0$ are left fixed at their best-fit values; $H_0$ is then set to be either $67.32$ or $74.03$ km/s/Mpc, the values reported by Planck 2018 (hereafter P18) and Riess, \textit{et al.} 2019 (hereafter R19) \cite{Riess:2019}, respectively. 

\begin{table}[h!]
\begin{center}
 \begin{tabular}{||c | c||}
 \hline
 Parameter & Best Fit \\ [0.5ex] 
 \hline\hline
 $H_0$ (km/s/Mpc) & 67.32 \\
 \hline
 $\Omega_c h^2$ & 0.12011 \\ 
 \hline
 $\Omega_b h^2$ & 0.022383 \\
 \hline
 $\tau$ & 0.0543 \\
 \hline
 $\ln(10^{10}A_s)$ & 3.0448 \\
 \hline
 $n_s$ & 0.96605 \\ [1ex] 
 \hline
\end{tabular}
\caption{Planck 2018 best-fit parameter values \cite{Aghanim:2018}. Future work will allow all of these parameters to vary. Note that $H_0$ is an inferred value from the fitted $100 \Theta_*$; they may be used interchangeably in the base parameter set.}
\label{tab:Params} 
\end{center}
\end{table}

For each $H_0$, we allow the DE equation of state to vary in the linear parametrization $w = w_0 + w_a(1-a)$. The distance, in terms of the ${\cal D}_{\rm JS}$, from each model to the Planck data can be summarized by two surfaces: $\mathcal{D}_\mathrm{JS}(w_0, w_a\:|\:h = 0.6732)$ and $\mathcal{D}_\mathrm{JS}(w_0, w_a\:|\:h = 0.7403)$. Our goal here is not to determine which value of $H_0$ is the ``correct'' one, but to investigate whether a model with a modified DE equation of state can shift the inferred $H_0$ closer to R19. Such a model would have a shorter distance (in the sense of the ${\cal D}_{\rm JS}$) to the Planck data and provide an alternative method of parameter estimation. A forthcoming full analysis will allow $H_0$ to vary along with the other parameters in Table \ref{tab:Params}.

\section{Results}\label{sec:Results} 

\begin{figure}[h!]
\includegraphics[width=\linewidth]{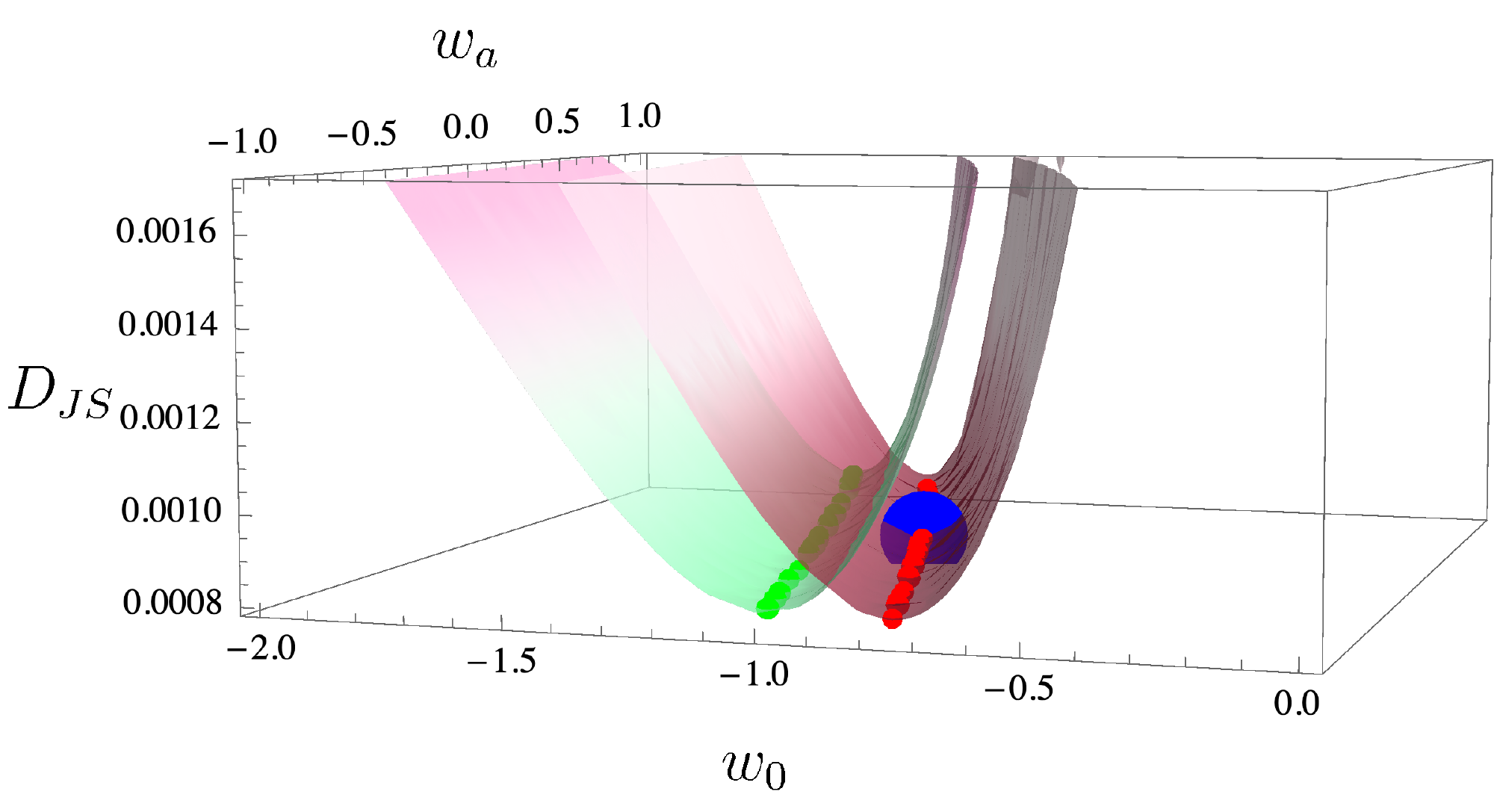}
\caption{${\cal D}_{\rm JS}$ surfaces for $\Lambda$CDM and R19 values of $H_0$ in red and green, respectively. $w_0$ and $w_a$ have been allowed to vary, but the other parameters were fixed at their best-fit values from Planck 2018 (see \autoref{tab:Params}). The red and green lines denote degeneracy in the model space; their distance from the Planck data is very nearly the same. The blue dot represents the $\Lambda$CDM model.}
\label{fig:w0wasurfaces}
\end{figure}

\begin{figure}[h!]
\includegraphics[width=\linewidth]{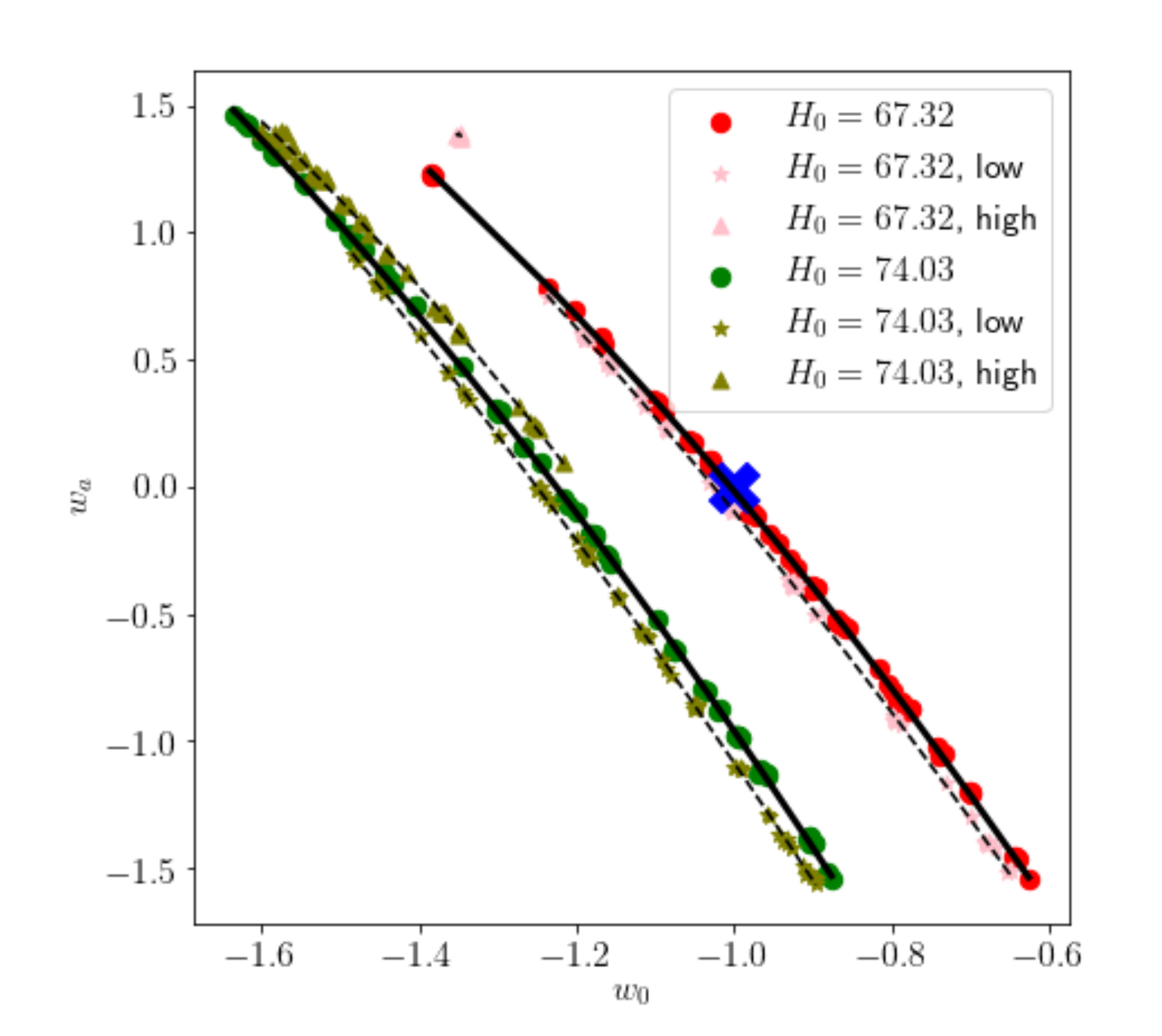}
\caption{Lines of degeneracy from Fig. \ref{fig:w0wasurfaces} projected into the $w_0-w_a$ plane and fit with a second-order polynomial, $w_a = c_0 w_0^2 + c_1 w_0 + c_2$. Here, $[c_0, c_1, c_2] = [-1.061,  -5.792, -4.750]$ and $[-1.013, -6.518,  -6.467]$ for P18 (right curve) and R19 (left curve) values of $H_0$, respectively. The solid lines represent the best fit to the acoustic power spectrum of the CMB, while the dashed lines represent fits to the high and low errors of the the CMB acoustic power spectrum. The finite lengths of these lines indicate a finite $w_0$-$w_a$ parameter space. The blue x represents the location of the $\Lambda$-CDM model.}
\label{fig:w0waValley}
\end{figure}

\autoref{fig:w0wasurfaces} shows the ${\cal D}_{\rm JS}$ as a function of $w_0$ and $w_a$ for both the P18 (right surface) and R19 (left surface) values of $H_0$. Both surfaces display a valley running along a degenerate minimum curve where ${\cal D}_{\rm JS}\simeq 8.442\times 10^{-4}$. Given the interpretation of ${\cal D}_{\rm JS}$ as a metric between the two distributions, its minimum represents the preferred parameter data set. On the $H_0 = 67.32$ surface, the $\Lambda$CDM model is identified with a larger blue point.  Figure  \ref{fig:w0waValley} shows the degenerate curves along the ${\cal D}_{\rm JS}$ valley projected onto the $(w_0,w_a)$ plane and fitted with a second-order polynomial. The dashed lines represent the error of the ${\cal D}_{\rm JS}$ minima, produced using the high and low errors in the measured values of the CMB TT acoustic power spectrum. Since the errors for each value of $H_0$ are small, the two curves are clearly distinguishable and do not overlap. Therefore, once one of the two parameters in the DE equation of state for this model is known, this approach allows for the determination of the other, thus breaking the degeneracy in a predictive way. In future work, we plan to apply a modified Fisher information matrix approach, adapted to  ${\cal D}_{\rm JS}$ used here and applicable to many variables, in order to address the issue of correlated data points and potential errors in parameter prediction. In the \hyperref[sec:Appendix]{Appendix}, we sketch the initial steps toward this formalism for a single variable.

\begin{figure}[h!]
\includegraphics[width=\linewidth]{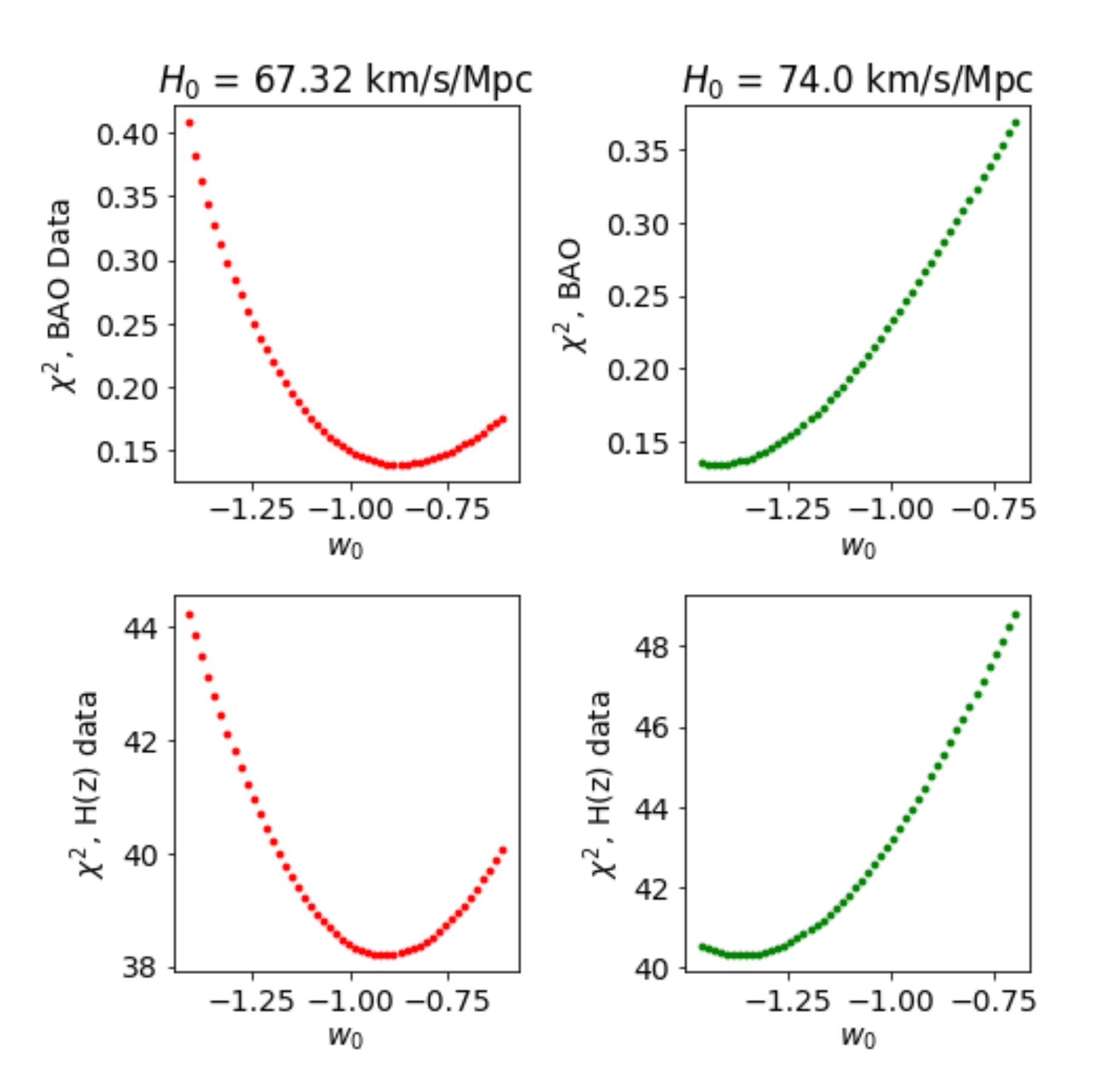}
\caption{For models along the curves of degeneracy found in \hyperref[fig:w0wasurfaces]{Fig. 2}, the $\chi^2$ is plotted as a function of $w_0$ for (top) the model's prediction of the BAO $D_V(z) / r_s(z_\mathrm{dec})$ and the measured data, and (bottom) the model's prediction for $H(z)$ and the measured data. The red and green curves come from using the P18 value and R19 value of $H_0$, respectively.}
\label{fig:BAO_Hz_chiSq}
\end{figure}

We also consider that data for the Hubble parameter as a function of redshift, $H(z)$---along with the BAO volume-averaged effective distance ratio $D_V(z) / r_s(z_\mathrm{dec})$---could break the degeneracy in this model space compared to using the CMB data only. While in future work we will include these data sets \textit{a priori} in the minimization of the ${\cal D}_{\rm JS}$ for a model, we use them here to examine their effect on the degeneracy found using only the Planck data. Taking pairs of values of $\omega_0$ and $\omega_a$ along the degenerate curves, we use $\chi^2$ to fit $38$ measurements of $H(z)$, compiled from \cite{Anagnostopoulos:2017} and references therein, and $12$ measurements of the BAO data, compiled from \cite{Cheng:2014, Haridasu:2017}. The results for $\chi^2(w_0)$ for each $H_0$ and both data sets are shown in \hyperref[fig:BAO_Hz_chiSq]{Fig. 4}.    

We find that including the additional late-time data sets does not reduce the degeneracy in a clear way.  \autoref{tab:Results} reports the values of $w_0$ and $w_a$ that minimize the $\chi^2$ for the BAO and $H(z)$ data, for both the P18 and R19 $H_0$.  Neither the BAO nor the \textit{H(z)} data show a significant difference between the $\chi^2$ for each $H_0$, indicating that the addition of these data sets would allow either $H_0$ value.

\begin{table}[h!]
\begin{center}
 \begin{tabular}{||c | c | c | c | c ||}
 \hline
 Data & $H_0$ (km/s/Mpc) & $w_0$ & $w_a$ & $\chi^2$ \\ [0.5ex] 
 \hline\hline
 BAO & P18 & $\:-0.87\:$ & $\:-0.51\:$ & $\:0.139\:$\\
 \hline
 BAO & R19 & $\:-1.41\:$ & $\:0.73\:$ & $\:0.134\:$\\ 
 \hline
 $H(z)$ & P18 & $\:-0.90\:$ &  $\:-0.38\:$ & $\:38.2 \:$\\
 \hline
 $H(z)$ & R19 & $\:-1.37\:$ & $\:0.55\:$ & $\:40.3\:$\\[1ex] 
 \hline
\end{tabular}
\caption{Values of $w_0$ and $w_a$ along the curves of degeneracy from \hyperref[fig:w0wasurfaces]{Fig. 2} that minimize the $\chi^2$ to the BAO and $H(z)$ data sets.}
\label{tab:Results} 
\end{center}
\end{table}

\vspace{-0.25cm}
\section{Concluding Remarks}\label{sec:Conclusions}
In this work, we introduced a new method to estimate cosmological parameters based on the Jensen-Shannon divergence ${\cal D}_{\rm JS}$ of information theory, inspired by the configurational entropy approach proposed in Ref. \cite{Gleiser:2012}. As a first application, we examined here the current tension in the value of the expansion rate $H_0$, comparing the extended $\Lambda$CDM temperature anisotropy spectrum for models with dynamic DE parametrized in $(w_0, w_a)$ space with the Planck 2018 temperature anisotropy data. For both values of $H_0$, we found that there are curves of degeneracy in the $(w_0,w_a)$ plane, characterized as nearly indistinguishable minima of the ${\cal D}_{\rm JS}(w_0,w_a)$ surface. However, the two curves are along different values of the model parameter pair $(\omega_0,~\omega_a)$, allowing for degeneracy breaking, indicating how our method could be used for the potential resolution of the $H_0$ tension, once one of the two parameters is known. (We expect that similar degeneracy lines would be found using the Bayesian approach.) Extending our analysis to include $H(z)$ and BAO at different redshifts, we found that the extended data do not lift the degeneracy. Taking our limited parameter exploration of this work at face value, our results indicate that a possible resolution may indeed come from early-universe modifications of the standard cosmological model.

In a forthcoming paper, we plan to extend this analysis by running a MCMC to minimize the ${\cal D}_{\rm JS}$ in the full seven-parameter space, and to include data from BAO, \textit{H(z)}, and the Planck polarization and temperature-polarization cross-correlation power spectra in our analysis. We expect this more complete approach to change the results plotted in Fig. \ref{fig:w0waValley}, which should be considered our method's first illustrative example. Our current results warrant further investigation of ${\cal D}_{\rm JS}$ as an alternative and transparent method of cosmological parameter estimation.
A more complete study will allow us to directly compare parameter confidence intervals from our information-based analysis to others reported in the literature. Work along these lines is currently in progress.

\section{acknowledgments}
 M.G. and M.S. were supported in part by a Department of Energy Grant No. DE-SC0010386 during the early stages of this collaboration, when M.S. was also supported by a Hull doctoral fellowship.

\pagebreak
\appendix
\section{APPENDIX\\
 Error estimation using an adapted inverse Fisher Information Matrix}\label{sec:Appendix}

Traditional model parameter estimation techniques use the inverse Fisher Information Matrix (FIM) to bound errors on the parameters from the Cramer-Rao inequality. We propose to adapt the FIM approach to our ${\cal D}_{\rm JS}$ approach. First, recall that for a set of $M$ model parameters $\Lambda = (\lambda_1,\lambda_2,...,\lambda_M)$ and a likelihood function $L(x_a;\Lambda)$, where $x_a$ is a data point belonging to a set of $O$ observables $\mathcal{O}=(x_1,x_2,...,x_O)$, the $M\times M$ FIM is 
\begin{equation}
\mathcal{F}_{\alpha\beta} = \langle \frac{ {\partial^2\mathcal{L}}} {{\partial \lambda_{\alpha}\partial \lambda_{\beta}} } \rangle,
\end{equation}
\noindent where $\mathcal{L}\equiv -\ln L$. Simplifying to a single model parameter $\lambda$, the Fisher matrix $\mathcal{F}$ is given by
\begin{equation}\label{eq:Fisher}
\mathcal{F} = \frac{1}{2}\frac{\partial^2 \mathcal{L}}{\partial \lambda ^2}|_{\lambda = \bar{\lambda}},
\end{equation}
\noindent evaluated at the fiducial value $\overline{\lambda}$ (the best guess). We may evaluate similarly for $\mathcal{D}_{\rm JS}(\lambda)$,
by expanding about $\lambda-\bar{\lambda}$,
\begin{equation}\label{eqn:exp}
\djs \simeq \djs (\bar{\lambda}) + \frac{\partial \djs}{\partial \lambda} |_{\lambda = \bar{\lambda}} (\lambda-\overline{\lambda})  + \frac{1}{2}\frac{\partial^2 \mathcal{D}_{\rm JS}}{\partial \lambda ^2}|_{\lambda = \bar{\lambda}}(\lambda-\overline{\lambda})^2,
\end{equation}
\noindent where we identify 
\begin{equation}
 \mathcal{F} \equiv \frac{1}{2}\frac{\partial^2 \mathcal{D}_{JS}}{\partial \lambda ^2}|_{\lambda = \bar{\lambda}}.
\end{equation}
The zeroth-order term vanishes as does the first-order term, since the slope of the $\djs$ curve at $\bar \lambda$ vanishes.

Using the definition of $\djs$ in \autoref{eq:JSD}, we write $\djs$ as a function of the data, $\clobs$, and the parameter-dependent model spectrum $C_l(\lambda)$ as, 

\begin{equation}
\begin{split}
\djs \left( \{ \lambda \} \right) & = -\frac{1}{2}\sum_l C_l^{obs} \ln\left( \frac{\frac{1}{2}(C_l^{obs}+
C_l(\lambda))}{C_l^{obs}}\right) \\
& -\frac{1}{2}\sum_l C_l(\lambda) \ln\left( \frac{\frac{1}{2}(C_l^{obs}+C_l(\lambda))}{C_l(\lambda)}\right).
\end{split}
\end{equation}
\noindent Taking the derivatives we obtain,
\begin{equation}
\mathcal{F} = -\frac{1}{2} \sum_l \frac{\partial ^2 C_l}{\partial \lambda^2} log \left(\frac{\rl}{C_l} \right).
\end{equation}
 \noindent Generalizing to include many parameters $\{\lambda_{\alpha}\}$ gives 
\begin{equation}
 \mathcal{F}_{\alpha\beta} = -\frac{1}{2} \sum_l \frac{\partial ^2 C_l}{\partial \lambda_\alpha \partial \lambda_\beta} \ln \left(\frac{\rl}{C_l} \right).
 \end{equation}
\noindent Since $\mathcal{F}$ is a symmetric matrix, the $1\sigma$ error on the parameters is
 \begin{equation}
 \sigma_{\alpha} \geq \frac{1}{\sqrt{\frac{1}{2} \mathcal{F}_{\alpha\alpha}}}.
 \end{equation}

\end{document}